\documentclass[ aps,%
12pt,%
floatfix,%
final,%
notitlepage,%
oneside,%
onecolumn,%
nobibnotes,%
nofootinbib,%
superscriptaddress,%
showpacs,%
noshowpacs,%
centertags]%
{revtex4}

\usepackage{amssymb,amsmath}
\usepackage{eufrak}
\usepackage{mathtext}
\usepackage[cp866]{inputenc}
\usepackage[T2A]{fontenc}
\usepackage[russian,english]{babel}
\usepackage{graphicx}
\usepackage{hyperref}

\newcommand{\QQ}{Q\bar Q}
\newcommand{\GeV}{\mbox{GeV}}
\newcommand{\mub}{\mu\mbox{b}}
\newcommand{\OS}{\langle O_S\rangle}
\newcommand{\OP}{\langle O_P\rangle}

\begin{document}
\title{Production of $\eta_Q$ meson at LHC}
\author{A.K. Likhoded}
\email{anatolii.likhoded@ihep.ru}
\affiliation{Institute for High Energy Physics, Protvino, Russia}
\affiliation{Moscow Institute of Physics and Technology, Dolgoprudny, Russia}

\author{A.V. Luchinsky}
\email{alexey.luchinsky@ihep.ru}
\affiliation{Institute for High Energy Physics, Protvino, Russia}
\affiliation{SSC RF ITEP of NRC "Kurchatov Institute"}

\author{S.V. Poslavsky}
\email{stvlpos@mail.ru}
\affiliation{Institute for High Energy Physics, Protvino, Russia}
\affiliation{SSC RF ITEP of NRC "Kurchatov Institute"}

\begin{abstract}
The paper is devoted to theoretical consideration of inclusive $\eta_{c,b}$ meson production at LHC. It is shown that existing experimental data on $\eta_c$ meson production at LHCb detector can be described in the framework of NRQCD formalism and color-singlet component with phenomenological value of matrix element $|R(0)|^2$ gives main contribution. Using this model we present theoretical prediction for integrated cross sections and transverse momentum distributions for inclusive $\eta_c$ production at other LHC detectors. The case of $\eta_b$ meson production at LHC is also considered.
\end{abstract}

\maketitle

\section{Introduction}

% new variant [AK]

Heavy quarkonia production can be described as production of heavy quark-antiquark pair $\QQ$ with small relative momentum in initial partons' interaction and subsequent projection of this pair to hadronic states with suitable quantum numbers. According to NRQCD \cite{Bodwin:1994jh} in addition to colour singlet component (CS) one has to take also into account contributions of colour-octet components (CO) with unknown nonperturbative long-distance matrix elements (LDME), that are treated as free parameters. In NRQCD the cross section of heavy quarkonium production is written as expansion over the small relative velocity $v$, so the number of unknown parameters is limited. As a result, one can hope that such approach could describe accurately momentum spectra of heavy quarkonia.

%Рождение тяжелого кваркония в последнее время представляется как жесткое рождение пары тяжелых кварков в столкновении начальных партонов и последующее проектирование $Q\bar Q$-системы с малым относительным импульсом на адронные состояния с нужными квантовыми числами. При этом, согласно NRQCD \cite{Bodwin:1994jh}, помимо состояния цветового синглета следует учесть и состояния цветового октета, непертурбативные матричные элементы которых неизвестны и являются свободными параметрами. NRQCD, понимаемая в случае тяжелого кваркония как ряд по степеням относительной скорости $v$, ограничивает число неизвестных параметров, что позволяет надеяться на описание импульсных спектров тяжелых кваркониев.

Unfortunately, currently NRQCD does not describe these spectra well enough, since one more problem exists. Usually integrated over transverse momenta distribution functions of initial partons  are used in calculations. There is one more approach, where this drawback is removed: so called $k_T$ factorization \cite{Hagler:2000dd, Hagler:2000eu,Kniehl:2006sk}. In this approach transverse momentum dependence of partonic distribution functions is taken into account. Unfotunatelly, this technique is a little bit ambiguous, so more intensive study is required.

%К сожалению, в настоящее время нельзя сказать, что NRQCD корректно описывает распределения по поперечному импульсу. Дело в том, что есть еще одна неопределенность в описании спектра, связанная с тем обстоятельством, что при вычислении, как правило, используются проингерированные по поперечному импульсу распределения начальных партонов. Существует подход, в этот недостаток снят и учтена зависимость функции распределения партона от его поперечного импульса ($k_T$-факторизация \cite{Hagler:2000dd, Hagler:2000eu,Kniehl:2006sk}). К сожалению, в этом подходе не все однозначно и требуется дальнейшее уточнение.

In our previous works \cite{Likhoded:2012hw, Likhoded:2013aya, Likhoded:2014kfa}, devoted to $P$-wave quarkonia states production in hadronic interaction, it was shown that in high transverse momentum region ($p_T>5\,\GeV$) one can use  $O(\alpha_s^3)$ approximation to describe $p_T$ spectrum of final qiuarkonium, while relative contributions of CS and CO states can be determined from analysis of $\chi_2/\chi_1$ ratio. In this model $p_T$ distribution is caused by emission of additional gluon in the initial state. The same effect also makes possible the production  of axial meson, that is forbidden in identical gluons' interaction by Lanau-Yang theorem. The only opened question in this approach is the enhance of CS contribution (or, in other words, derivative of the quarkonium's wave function in the origin). In the work \cite{Jia:2014jfa}, on the other hand, it is argued that NLO corrections allows one to reproduce experimental data using CS LDME determined from $\chi_{c2}$ decay width. Recently Collaboration LHCb has published new data on $\eta_c$ meson production, so one can test the availablilty of mentioned method in this case. Below we present the results of analysis of new LHCb data.

\section{Inclusive $\eta_Q$ Production}

%Recently a bunch of theoretical and experimental works devoted to inclusive production of different $S$- and $P$-wave heavy quarkonia states at LHC have appeared. For example, in works \cite{QQQ} production of $J/\psi$ meson is considered and it is shown that contributions of colour octet (CO) components are large. Analysis, presented in \cite{Likhoded:2014kfa}, on the contrary, shows that in the case of $\chi_c$ meson production colour singlet (CS) components give main contributions with the corresponding matrix element $|R_{cc}'(0)|^2$ about 3 times larger than the phenomenological value, obtained from the width of $\chi_{c2}$ meson or from potential models. 
Recently LHCb collaboration has published new experimental work devoted to inclusive $\eta_c$ meson production \cite{Aaij:2014bga}, so it would be extremely interesting to study this process theoretically. This is the topic of the current note.

\begin{figure}
\includegraphics[width=\textwidth]{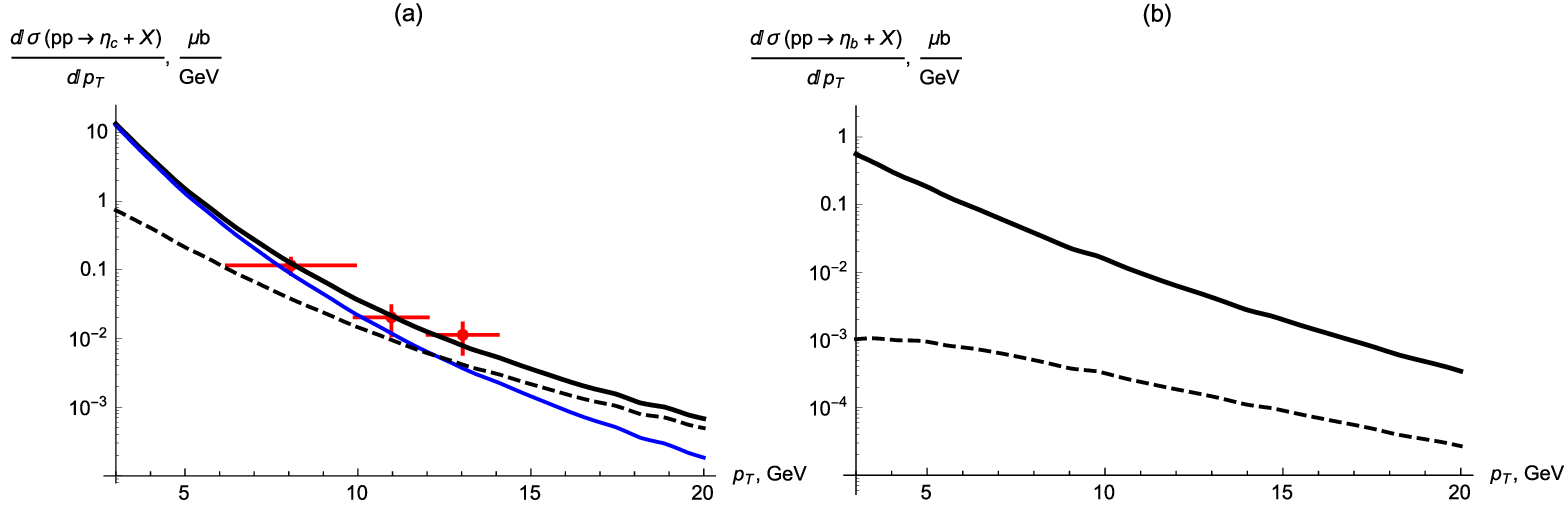}
\caption{
Transverse momentum distribution in $pp\to\eta_c+X$ (left figure) and $pp\to\eta_b+X$ (right figure) at LHCb. Solid black, blue, and dashed black lines in the left figure correspond to total, CS and $S$-wave CO cross actions respectively, experimental data are taken from \cite{Aaij:2014bga}.
}
\label{fig}
\end{figure}

According to NRQCD scaling rules \cite{Bodwin:1994jh} (see also \cite{Biswal:2010xk}) CS, as well as $S$- and $P$-wave CO components should give main contributions to the cross section of the process under consideration:
\begin{eqnarray}
\frac{d\hat\sigma(gg\to\eta_Q g)}{dp_T} &=&
|R(0)|^2 \frac{d\hat\sigma(gg\to Q\bar Q[^1S_0^{[1]}]g)}{dp_T} +
 \OS \frac{d\hat\sigma(gg\to Q\bar Q[^3S_1^{[8]}]g)}{dp_T} + 
 \nonumber \\ &&
  \OP \frac{d\hat\sigma(gg\to Q\bar Q[^1P_1^{[8]}]g)}{dp_T},
\end{eqnarray}
where we use results of the work \cite{Meijer:2007eb} for the expressions of the hard cross sections, $R(0)$ is the value of the heavy quarkonum wave function in the CS state at the origin, and the following notations for CO LDME are used \cite{Meijer:2007eb}:
\begin{eqnarray}
\OS &=& \langle R_{\eta_Q} [^3S_1^{8)}]\rangle = \frac{\pi}{6}\langle 0 | \mathcal{O}_8^{\eta_Q}[^3S_1]  | 0 \rangle,
\\
\OP &=& \langle R_{\eta_Q}[^1P_1^{(8)}] \rangle = \frac{\pi}{18} \langle 0 | \mathcal{O}_8^{\eta_Q}[^1P_1]  | 0 \rangle.
\end{eqnarray}
Color singlet LDME can be determined potential models \cite{Munz:1996hb,Ebert:2003mu,Anisovich:2005jp,Wang:2009er,Li:2009nr,Hwang:2010iq} or  leptonic width of $J/\psi$ meson:
\begin{eqnarray}
|R_{cc}(0)|^2 &=& \frac{M^2 \Gamma(J/\psi\to e^+e^-)}{4\alpha^2 e_c^2} \approx 0.58\,\GeV^3.
\label{eq:R0}
\end{eqnarray}
As for octet LDME, our analysis shows, that experimental data \cite{Aaij:2014bga} can be described taking into account only contribution of $S$-wave CO component with the matrix elemet
\begin{eqnarray}
1.5\times 10^{-3}\,\GeV^3 &<& \OS <  5.3\times 10^{-3}\,\GeV^3,
\label{eq:OS}
\end{eqnarray}
that corresponds to
\begin{eqnarray}
2.9\times 10^{-3}\,\GeV^3 &<& \langle O^{\eta_c}_8[^3S_1]\rangle < 1.0\times 10^{-2}\,\GeV^3.
\label{eq:OS2}
\end{eqnarray}
The contribution of $P$-wave CO competent is strongly suppressed. Using presented above parameter values it is easy to obtain the following cross section of $\eta_c$ meson production at LHCb (the cut $p_T>6.5\,\GeV$ is imposed on the transverse momentum of the final charmonium):
\begin{eqnarray}
\sigma_{\mathrm{LHCb}}^{\mathrm{th}}[pp\to\eta_c(1S)+X] &=& 0.58\,\mub.
\label{eq:sigma_th}
\end{eqnarray}
The contributions of CS and $P$-wave CO components are about 70\% and 30\% respectively. The value \eqref{eq:sigma_th} is in good agreement with experimental result
\begin{eqnarray}
\sigma_{\mathrm{LHCb}}^{\mathrm{exp}}[pp\to\eta_c(1S)+X] &=& 0.52\pm0.11\pm0.09\pm0.08\,\mub.
\end{eqnarray}
It can be seen from Fig.\ref{fig}a, that transverse momentum distribution are also in good agreement with experimental data. It should be noted, that presented above values of the LDMEs are strongly correlated. For example, increasing the value of the CS matrix element up to the value $|R(0)|^2\approx 0.8\,\GeV^3$ one can exclude the CO states completely.

It is interesting to compare our results with the results presented in \cite{Biswal:2010xk}. According to Case-I fit of the last paper (see also discussion in \cite{Biswal:2010xk}) experimental data can be described by CS contributions only and $S$-wave CO components increase the cross section by about two orders of magnitude. The value of corresponding LDME in \cite{Biswal:2010xk} is
\begin{eqnarray}
\langle O^{\eta_c}_8[^1S_0] \rangle &=& (6.6\pm1.5)\times 10^{-2}\,\GeV^3,
\end{eqnarray}
that is significantly larger than in \eqref{eq:OS2}. The reason for this discrepancy could be caused by the fact, that CO matrix elements in \cite{Biswal:2010xk} were obtained from the fit of $J/\psi$ production cross section at CDF \cite{Cho:1995ce}, where CO components give main contribution.

We have mentioned above, that of production of other heavy quarkonia states is considered the situation is quite different: either contribution of CO components is dominant or the value of singlet LDME is anomalously large. So we think that independent check of LHCb results at other detectors would be desirable. Using matrix elements from eq.\eqref{eq:R0}, \eqref{eq:OS} it is easy to obtain the values of the $pp\to\eta_c+X$ cross sections at CMS and ATLAS detectors, presented in table \ref{tab}.

In the framework of the same model one can obtain also theoretical predictions for $\eta_b$ meson production at LHC. The CS matrix element is this case equals
\begin{eqnarray}
|R_{bb}(0)|^2 &\approx& 5.3\,\GeV^3,
\end{eqnarray}
while CO matrix element can be estimated using dimensional arguments, as it was done in paper \cite{Likhoded:2012hw}:
\begin{eqnarray}
\langle O_S^{\eta_b} \rangle &=& \frac{M_{\eta_c}^2}{M_{\eta_b}^2}\frac{|R_{bb}(0)|^2}{|R_{cc}(0)|^2}\langle O_S^{\eta_c}\rangle
\approx 0.01\,\GeV^3.
\end{eqnarray}
Resulting cross sections can be found in table \ref{tab} and transverse momentum distribution is shown in Fig.\ref{fig}b. It turns out that for all detectors the contribution of CO component is about 2\%.

\begin{table}
\begin{tabular}{|c|c|c|}
\hline
exp & $\eta_c$ & $\eta_b$ \\
\hline
LHCb & 0.58   & 0.17 \\ 
CMS  &  0.85   & 0.28  \\
ATLAS & 0.63  & 0.22  \\
\hline
\end{tabular}
\caption{
Cross sections of $\eta_{c,b}$ meson production (in $\mub$) at LHC. In all cases the cut $p_T>6.5\,\GeV$ is imposed on the transverse momentum of final quarkonium.}
\label{tab}
\end{table}

\section{Conclusion}

The paper is devoted to theoretical analysis of inclusive heavy quarkonia $\eta_{c,b}$ production at LHC. It is shown that in the case of $\eta_c$ meson experimental data, obtained by LHCb collaboration, can be described in the framework of NRQCD model with phenomenological value of colour singlet matrix element and CS component gives main contribution. Using the same model we obtain predictions for cross sections and transverse momentum distributions at other LHCb detectors and $pp\to\eta_b+X$ reaction.

The work was financially supported by RFBR (grant \#14-02-00096 A) and grant of SAEC "Rosatom" and Helmholtz Association.

%\bibliographystyle{utphys}
%\bibliographystyle{apsrev}
%\bibliography{refs}

\begin{thebibliography}{18}
\expandafter\ifx\csname natexlab\endcsname\relax\def\natexlab#1{#1}\fi
\expandafter\ifx\csname bibnamefont\endcsname\relax
  \def\bibnamefont#1{#1}\fi
\expandafter\ifx\csname bibfnamefont\endcsname\relax
  \def\bibfnamefont#1{#1}\fi
\expandafter\ifx\csname citenamefont\endcsname\relax
  \def\citenamefont#1{#1}\fi
\expandafter\ifx\csname url\endcsname\relax
  \def\url#1{\texttt{#1}}\fi
\expandafter\ifx\csname urlprefix\endcsname\relax\def\urlprefix{URL }\fi
\providecommand{\bibinfo}[2]{#2}
\providecommand{\eprint}[2][]{\url{#2}}

\bibitem[{\citenamefont{Bodwin et~al.}(1995)\citenamefont{Bodwin, Braaten, and
  Lepage}}]{Bodwin:1994jh}
\bibinfo{author}{\bibfnamefont{G.~T.} \bibnamefont{Bodwin}},
  \bibinfo{author}{\bibfnamefont{E.}~\bibnamefont{Braaten}}, \bibnamefont{and}
  \bibinfo{author}{\bibfnamefont{G.~P.} \bibnamefont{Lepage}},
  \bibinfo{journal}{Phys.Rev.} \textbf{\bibinfo{volume}{D51}},
  \bibinfo{pages}{1125} (\bibinfo{year}{1995}), \eprint{hep-ph/9407339}.

\bibitem[{\citenamefont{Hagler et~al.}(2001{\natexlab{a}})\citenamefont{Hagler,
  Kirschner, Schafer, Szymanowski, and Teryaev}}]{Hagler:2000dd}
\bibinfo{author}{\bibfnamefont{P.}~\bibnamefont{Hagler}},
  \bibinfo{author}{\bibfnamefont{R.}~\bibnamefont{Kirschner}},
  \bibinfo{author}{\bibfnamefont{A.}~\bibnamefont{Schafer}},
  \bibinfo{author}{\bibfnamefont{L.}~\bibnamefont{Szymanowski}},
  \bibnamefont{and} \bibinfo{author}{\bibfnamefont{O.}~\bibnamefont{Teryaev}},
  \bibinfo{journal}{Phys.Rev.Lett.} \textbf{\bibinfo{volume}{86}},
  \bibinfo{pages}{1446} (\bibinfo{year}{2001}{\natexlab{a}}),
  \eprint{hep-ph/0004263}.

\bibitem[{\citenamefont{Hagler et~al.}(2001{\natexlab{b}})\citenamefont{Hagler,
  Kirschner, Schafer, Szymanowski, and Teryaev}}]{Hagler:2000eu}
\bibinfo{author}{\bibfnamefont{P.}~\bibnamefont{Hagler}},
  \bibinfo{author}{\bibfnamefont{R.}~\bibnamefont{Kirschner}},
  \bibinfo{author}{\bibfnamefont{A.}~\bibnamefont{Schafer}},
  \bibinfo{author}{\bibfnamefont{L.}~\bibnamefont{Szymanowski}},
  \bibnamefont{and} \bibinfo{author}{\bibfnamefont{O.}~\bibnamefont{Teryaev}},
  \bibinfo{journal}{Phys.Rev.} \textbf{\bibinfo{volume}{D63}},
  \bibinfo{pages}{077501} (\bibinfo{year}{2001}{\natexlab{b}}),
  \eprint{hep-ph/0008316}.

\bibitem[{\citenamefont{Kniehl et~al.}(2006)\citenamefont{Kniehl, Vasin, and
  Saleev}}]{Kniehl:2006sk}
\bibinfo{author}{\bibfnamefont{B.}~\bibnamefont{Kniehl}},
  \bibinfo{author}{\bibfnamefont{D.}~\bibnamefont{Vasin}}, \bibnamefont{and}
  \bibinfo{author}{\bibfnamefont{V.}~\bibnamefont{Saleev}},
  \bibinfo{journal}{Phys.Rev.} \textbf{\bibinfo{volume}{D73}},
  \bibinfo{pages}{074022} (\bibinfo{year}{2006}), \eprint{hep-ph/0602179}.

\bibitem[{\citenamefont{Likhoded et~al.}(2012)\citenamefont{Likhoded,
  Luchinsky, and Poslavsky}}]{Likhoded:2012hw}
\bibinfo{author}{\bibfnamefont{A.}~\bibnamefont{Likhoded}},
  \bibinfo{author}{\bibfnamefont{A.}~\bibnamefont{Luchinsky}},
  \bibnamefont{and}
  \bibinfo{author}{\bibfnamefont{S.}~\bibnamefont{Poslavsky}},
  \bibinfo{journal}{Phys.Rev.} \textbf{\bibinfo{volume}{D86}},
  \bibinfo{pages}{074027} (\bibinfo{year}{2012}), \eprint{arXiv:1203.4893}.

\bibitem[{\citenamefont{Likhoded et~al.}(2013)\citenamefont{Likhoded,
  Luchinsky, and Poslavsky}}]{Likhoded:2013aya}
\bibinfo{author}{\bibfnamefont{A.}~\bibnamefont{Likhoded}},
  \bibinfo{author}{\bibfnamefont{A.}~\bibnamefont{Luchinsky}},
  \bibnamefont{and} \bibinfo{author}{\bibfnamefont{S.}~\bibnamefont{Poslavsky}}
  (\bibinfo{year}{2013}), \eprint{arXiv:1305.2389 [hep-ph]}.

\bibitem[{\citenamefont{Likhoded et~al.}(2014)\citenamefont{Likhoded,
  Luchinsky, and Poslavsky}}]{Likhoded:2014kfa}
\bibinfo{author}{\bibfnamefont{A.}~\bibnamefont{Likhoded}},
  \bibinfo{author}{\bibfnamefont{A.}~\bibnamefont{Luchinsky}},
  \bibnamefont{and}
  \bibinfo{author}{\bibfnamefont{S.}~\bibnamefont{Poslavsky}},
  \bibinfo{journal}{Phys.Rev.} \textbf{\bibinfo{volume}{D90}},
  \bibinfo{pages}{074021} (\bibinfo{year}{2014}), \eprint{arXiv:1409.0693}.

\bibitem[{\citenamefont{Jia et~al.}(2014)\citenamefont{Jia, Yu, and
  Zhang}}]{Jia:2014jfa}
\bibinfo{author}{\bibfnamefont{L.}~\bibnamefont{Jia}},
  \bibinfo{author}{\bibfnamefont{L.}~\bibnamefont{Yu}}, \bibnamefont{and}
  \bibinfo{author}{\bibfnamefont{H.-F.} \bibnamefont{Zhang}}
  (\bibinfo{year}{2014}), \eprint{arXiv:1410.4032}.

\bibitem[{\citenamefont{Aaij et~al.}(2014)}]{Aaij:2014bga}
\bibinfo{author}{\bibfnamefont{R.}~\bibnamefont{Aaij}} \bibnamefont{et~al.}
  (\bibinfo{collaboration}{LHCb collaboration}) (\bibinfo{year}{2014}),
  \eprint{arXiv:1409.3612}.

\bibitem[{\citenamefont{Biswal and Sridhar}(2012)}]{Biswal:2010xk}
\bibinfo{author}{\bibfnamefont{S.~S.} \bibnamefont{Biswal}} \bibnamefont{and}
  \bibinfo{author}{\bibfnamefont{K.}~\bibnamefont{Sridhar}},
  \bibinfo{journal}{J.Phys.} \textbf{\bibinfo{volume}{G39}},
  \bibinfo{pages}{015008} (\bibinfo{year}{2012}), \eprint{arXiv:1007.5163}.

\bibitem[{\citenamefont{Meijer et~al.}(2008)\citenamefont{Meijer, Smith, and
  van Neerven}}]{Meijer:2007eb}
\bibinfo{author}{\bibfnamefont{M.}~\bibnamefont{Meijer}},
  \bibinfo{author}{\bibfnamefont{J.}~\bibnamefont{Smith}}, \bibnamefont{and}
  \bibinfo{author}{\bibfnamefont{W.}~\bibnamefont{van Neerven}},
  \bibinfo{journal}{Phys.Rev.} \textbf{\bibinfo{volume}{D77}},
  \bibinfo{pages}{034014} (\bibinfo{year}{2008}), \eprint{arXiv:0710.3090}.

\bibitem[{\citenamefont{Munz}(1996)}]{Munz:1996hb}
\bibinfo{author}{\bibfnamefont{C.~R.} \bibnamefont{Munz}},
  \bibinfo{journal}{Nucl.Phys.} \textbf{\bibinfo{volume}{A609}},
  \bibinfo{pages}{364} (\bibinfo{year}{1996}), \eprint{hep-ph/9601206}.

\bibitem[{\citenamefont{Ebert et~al.}(2003)\citenamefont{Ebert, Faustov, and
  Galkin}}]{Ebert:2003mu}
\bibinfo{author}{\bibfnamefont{D.}~\bibnamefont{Ebert}},
  \bibinfo{author}{\bibfnamefont{R.}~\bibnamefont{Faustov}}, \bibnamefont{and}
  \bibinfo{author}{\bibfnamefont{V.}~\bibnamefont{Galkin}},
  \bibinfo{journal}{Mod.Phys.Lett.} \textbf{\bibinfo{volume}{A18}},
  \bibinfo{pages}{601} (\bibinfo{year}{2003}), \eprint{hep-ph/0302044}.

\bibitem[{\citenamefont{Anisovich et~al.}(2007)\citenamefont{Anisovich, Dakhno,
  Matveev, Nikonov, and Sarantsev}}]{Anisovich:2005jp}
\bibinfo{author}{\bibfnamefont{V.}~\bibnamefont{Anisovich}},
  \bibinfo{author}{\bibfnamefont{L.}~\bibnamefont{Dakhno}},
  \bibinfo{author}{\bibfnamefont{M.}~\bibnamefont{Matveev}},
  \bibinfo{author}{\bibfnamefont{V.}~\bibnamefont{Nikonov}}, \bibnamefont{and}
  \bibinfo{author}{\bibfnamefont{A.}~\bibnamefont{Sarantsev}},
  \bibinfo{journal}{Phys.Atom.Nucl.} \textbf{\bibinfo{volume}{70}},
  \bibinfo{pages}{63} (\bibinfo{year}{2007}), \eprint{hep-ph/0510410}.

\bibitem[{\citenamefont{Wang}(2009)}]{Wang:2009er}
\bibinfo{author}{\bibfnamefont{G.-L.} \bibnamefont{Wang}},
  \bibinfo{journal}{Phys.Lett.} \textbf{\bibinfo{volume}{B674}},
  \bibinfo{pages}{172} (\bibinfo{year}{2009}), \eprint{arXiv:0904.1604}.

\bibitem[{\citenamefont{Li and Chao}(2009)}]{Li:2009nr}
\bibinfo{author}{\bibfnamefont{B.-Q.} \bibnamefont{Li}} \bibnamefont{and}
  \bibinfo{author}{\bibfnamefont{K.-T.} \bibnamefont{Chao}},
  \bibinfo{journal}{Commun.Theor.Phys.} \textbf{\bibinfo{volume}{52}},
  \bibinfo{pages}{653} (\bibinfo{year}{2009}), \eprint{arXiv:0909.1369}.

\bibitem[{\citenamefont{Hwang and Guo}(2010)}]{Hwang:2010iq}
\bibinfo{author}{\bibfnamefont{C.-W.} \bibnamefont{Hwang}} \bibnamefont{and}
  \bibinfo{author}{\bibfnamefont{R.-S.} \bibnamefont{Guo}},
  \bibinfo{journal}{Phys.Rev.} \textbf{\bibinfo{volume}{D82}},
  \bibinfo{pages}{034021} (\bibinfo{year}{2010}), \eprint{arXiv:1005.2811}.

\bibitem[{\citenamefont{Cho and Leibovich}(1996)}]{Cho:1995ce}
\bibinfo{author}{\bibfnamefont{P.~L.} \bibnamefont{Cho}} \bibnamefont{and}
  \bibinfo{author}{\bibfnamefont{A.~K.} \bibnamefont{Leibovich}},
  \bibinfo{journal}{Phys.Rev.} \textbf{\bibinfo{volume}{D53}},
  \bibinfo{pages}{6203} (\bibinfo{year}{1996}), \eprint{hep-ph/9511315}.

\end{thebibliography}

\end{document}